\documentclass{svproc}
\usepackage[a4paper, total={6in, 8in}]{geometry}
\usepackage{url}

\usepackage{times}
\usepackage{dblfloatfix}  
\usepackage{float}
\usepackage{caption}
\usepackage{amsmath,amssymb,amsfonts}
\usepackage{graphicx}
\usepackage{textcomp}
\usepackage{xcolor,subfigure}
\usepackage{dblfloatfix}  
\usepackage{float}
\usepackage{bbding}
\begin{document}
\mainmatter              
	\title{Detect \& Reject for Transferability of Black-box Adversarial Attacks  Against Network Intrusion Detection Systems}
\titlerunning{Adversarial Attacks against Intrusion Detection Systems}   
%
	\author{Islam Debicha\inst{1,2}
	\and Thibault Debatty\inst{1} 
	\and Jean-Michel Dricot\inst{2} 
	\and Wim Mees\inst{1} 
	\and Tayeb Kenaza\inst{3}}
	\authorrunning{Islam Debicha et al.} 
%
	\tocauthor{Islam Debicha, Thibault Debatty,	Jean-Michel Dricot, Wim Mees, and Tayeb Kenaza}
	\institute{Royal Military Academy, Rue Hobbema 8, 1000 Brussels, Belgium,\\
	  \email{\{thibault.debatty,wim.mees\}@rma.ac.be}
	\and
	Universit\'{e} libre de Bruxelles, Avenue Franklin Roosevelt 50, 1050 Brussels, Belgium, \\
	\email{debichasislam@gmail.com}, \email{jean-michel.dricot@ulb.be}
	\and
	\'{E}cole Militaire Polytechnique, Bordj El-Bahri 17, 16111 Algiers, Algeria,\\
	\email{ken.tayeb@gmail.com}
}

\maketitle              

\begin{abstract}
	In the last decade, the use of Machine Learning techniques in anomaly-based intrusion detection systems has seen much success. However, recent studies have shown that Machine learning in general and deep learning specifically are vulnerable to adversarial attacks where the attacker attempts to fool models by supplying deceptive input. Research in computer vision, where this vulnerability was first discovered, has shown that adversarial images designed to fool a specific model can deceive other machine learning models. In this paper, we investigate the transferability of adversarial network traffic against multiple machine learning-based intrusion detection systems. Furthermore, we analyze the robustness of the ensemble intrusion detection system, which is notorious for its better accuracy compared to a single model, against the transferability of adversarial attacks. Finally, we examine Detect \& Reject as a defensive mechanism to limit the effect of the transferability property of adversarial network traffic against machine learning-based intrusion detection systems.
\end{abstract}

\section{Introduction}

The computer and networking industry is becoming increasingly important due to their growing use in various fields, which in turn leads to an escalation in the occurrence of cyberattacks. As a result, security is becoming a key concern of any network architecture and an active research topic. Intrusion Detection Systems (IDS) are one of the solutions presented to enhance network security by analyzing the traffic to identify any suspicious activity.  

In order to detect intrusions, there are mainly two approaches: the first one is based on comparing traffic with a list of all known attack patterns, also called signature-based intrusion detection. Intuitively, this method gives excellent accuracy when dealing with known attacks, however, this type of detector is incapable of detecting zero-day attacks, which is essentially what motivates the use of the second type of intrusion detection techniques, called anomaly-based intrusion detection. The latter approach relies on modeling the normal behavior of network traffic and later examining new traffic against this baseline.

Anomaly-based intrusion detection has been extensively studied, with most research using Machine Learning (ML) techniques to create a trustworthy model of activity due to their high accuracy, including deep learning which is considered a state-of-the-art technique in this field. However, recent studies have shown that machine learning in general, and deep learning in particular, are vulnerable to adversarial attacks where the attacker seeks to fool the models by inserting slight but specially crafted distortions into the original input \cite{akhtar2018threat}. 

Research in the field of computer vision, where this vulnerability was first discovered, has shown that adversarial images designed to fool a specific model can, to some extent, fool other machine learning models \cite{papernot2016transferability}. This is known as the transferability property of adversarial attacks. By exploiting this property, an attacker can build a surrogate intrusion detection system, create adversarial traffic for that detector, and then attack another intrusion detection system without even knowing the internal architecture of that detector, leading to a black-box attack.

To avoid this kind of vulnerability, we are conducting this research and the following are our contributions in this paper:
\begin{itemize}
	\item 
	To the best of our knowledge, this is the first study to examine the transferability of adversarial network traffic between multiple anomaly-based intrusion detection systems with different machine learning techniques in black-box settings.
	\item 
	In addition, we construct an ensemble intrusion detection system to examine its robustness against the transferability property of adversarial attacks compared to single detectors.
	\item 
	Finally, we investigate the effectiveness of the Detect \& Reject method as a defensive mechanism to mitigate the effect of the transferability property of adversarial network traffic against machine learning-based intrusion detection systems.
\end{itemize}

\section{Background}
\label{sec:back}

\subsection{Related Work}

Studies have shown that incorporating machine learning techniques can help improving the performance of intrusion detection systems. Aslahi-Shahri et al. \cite{aslahi2016hybrid} have proposed and explained the implementation of an intrusion detection system based on a hybrid support vector machine (SVM) and genetic algorithm (GA) method. In \cite{debicha2020efficient}, the authors have proposed a novel evidential IDS based on Dempster-Shafer theory to take into account source reliability. Vinayakumar et al. \cite{vinayakumar2019deep} have proposed a highly scalable intrusion detection framework using deep neural networks (DNNs) after a comprehensive evaluation of their performance against classical machine learning classifiers. Alamiedy et al.\cite{alamiedy2019anomaly} proposed an improved anomaly-based IDS model based on a multi-objective gray wolf optimization (GWO) algorithm, in which GWO is used as a feature selection technique. Ghanem et al.\cite{ghanem2020metaheuristic} have proposed a cyber-intrusion detecting system classification with MLP trained by a hybrid metaheuristic algorithm and feature selection based on multi-objective wrapper method. In \cite{ghanem2020efficient}, the authors proposed a new binary classification model for intrusion detection, based on hybridization of Artificial Bee Colony algorithm (ABC) and Dragonfly algorithm (DA) for training an artificial neural network (ANN) in order to increase the classification accuracy rate for malicious and non-malicious traffic in networks. Nevertheless, we noticed that little or no attention was paid to the effect of adversarial attacks when proposing these solutions.

Szegedy et al.\cite{szegedy2013intriguing} was the first work to report the vulnerability of DNN to adversarial samples where they introduced imperceptible adversarial perturbations to handwritten digits images and succeeded to fool the DNN model with high confidence. This discovery has prompted a number of studies in the computer vision community, where several attacks and defenses have been proposed \cite{goodfellow2014explaining,madry2017towards,grosse2017statistical}. There are some works \cite{papernot2016transferability,xie2019improving,lu2020enhancing} dealing with the transferability of adversarial attacks. However, these studies were specifically designed for the field of image classification, where this particular vulnerability was first detected.

In recent studies, the effect of adversarial attacks on intrusion detection systems has been investigated. Wang \cite{wang2018deep} Inspected the performance of state-of-the-art attack algorithms against deep learning-based intrusion detection on the NSL-KDD dataset. Pawlicki et al.\cite{pawlicki2020defending} evaluated the possibility of deteriorating the performance of a well-optimized intrusion detection algorithm at test time by generating adversarial attacks and then offers a way to detect those attacks.

Through our literature review, we did not find any studies on the transferability of adversarial network traffic between multiple anomaly-based intrusion detection systems with different machine learning techniques in black-box settings. Hence, we propose to conduct this study in order to fill this gap. Notice that in our recent work \cite{debicha2021adversarial}, we investigated the effectiveness of adversarial training as a defense for intrusion detection systems against these attacks.
%

\subsection{ Adversarial Attacks}


Despite their considerable success in achieving high accuracy, machine learning algorithms in general and deep learning, in particular, have proven vulnerable to adversarial attacks, where crafting an instance with small intentional perturbations can lead a machine learning model to make an erroneous prediction \cite{szegedy2013intriguing}.  

The idea of generating adversarial examples is quite intuitive. It can be seen as the inverse process of gradient descent where, for a given input $x$ and its label $y$, one tries to find model parameters $\theta$ that maximize the accuracy of the model by minimizing the loss function $J$. On the other hand, the adversarial examples are generated in order to minimize the accuracy of the model. Given the parameters $\theta$, the loss function $J$ is differentiated with respect to the input data $x$ so as to find an instance $x'$, as close to $x$ as possible, that maximizes the loss function $J$.

One of the earliest and most popular adversarial attacks is called the Fast Gradient Sign Method (FGSM) \cite{goodfellow2014explaining}. This attack uses a factor $\epsilon$ to limit the amount of distortion in the original instance such that $\lVert x'-x\lVert< \epsilon$ . One can think of $\epsilon$ as the attack strength or the size of the introduced perturbation. 
An adversarial instance $x'$ is devised like following:
\begin{equation}\label{key}
	x'= x + \epsilon\nabla J_{x}(x,y,\theta)
\end{equation}

Projected Gradient Descent (PGD) \cite{madry2017towards} is another adversarial attack and is basically an iterative extension of FGSM applying the attack repeatedly. However, PGD initializes the instance, at each iteration, to a random point in the $\epsilon$-ball around the original input.

Adversarial attacks can be classified into two types based on the attacker's knowledge of the attacked system:
White-box setting where the attacker has full knowledge of the internal architecture of the attacked system.
Black-box setting where the attacker has no knowledge of the internal architecture of the attacked system.
In this paper, we create adversarial network traffic records against a DNN-based IDS in a white-box setting, and then the rest of the experiments are conducted in a black-box setting where we use these adversarial instances to attack the other ML-based IDSs without having access to their internal architecture or training data.

\section{Proposed Approach}
\label{sec:ea}
Previous work has shown that the accuracy of a DNN-based IDS can be significantly reduced when exposed to adversarial attacks \cite{wang2018deep,pawlicki2020defending,debicha2021adversarial}. In this section, we construct a DNN-based IDS and five other ML-based IDSs to examine whether the same adversarial instances designed for a DNN-based IDS can be transferred to other ML-based IDSs, which are trained on a different data set, without knowing anything about their internal architectures. We also build an ensemble intrusion detection by clustering the five ML-based IDSs to study whether having multiple classifiers voting prediction can be a defense against the transferability property of adversarial attacks. Finally, we implement the Detect \& Reject method as a defense mechanism for the intrusion detection system and evaluate its robustness to transferable adversarial examples.    
\subsection{NSL-KDD Dataset}

Many research papers in the intrusion detection community have used the NSL-KDD dataset to demonstrate the performance of their proposed approaches and this is very useful as it creates a common basis for comparison between these approaches. Published in 2009 by \cite{tavallaee2009detailed}, this dataset is an improvement of the well-known KDD-CUP'99 which has a fundamental problem of record redundancy which leads to the bias of classifiers towards frequent records. This problem has been solved in the NSL-KDD dataset by proposing a balanced version of KDD-CUP'99 by removing the redundancy, thus providing a more accurate comparative analysis of the performance of different proposed intrusion detection frameworks.

\begin{figure}
	\centering
	\includegraphics[width=0.7\linewidth]{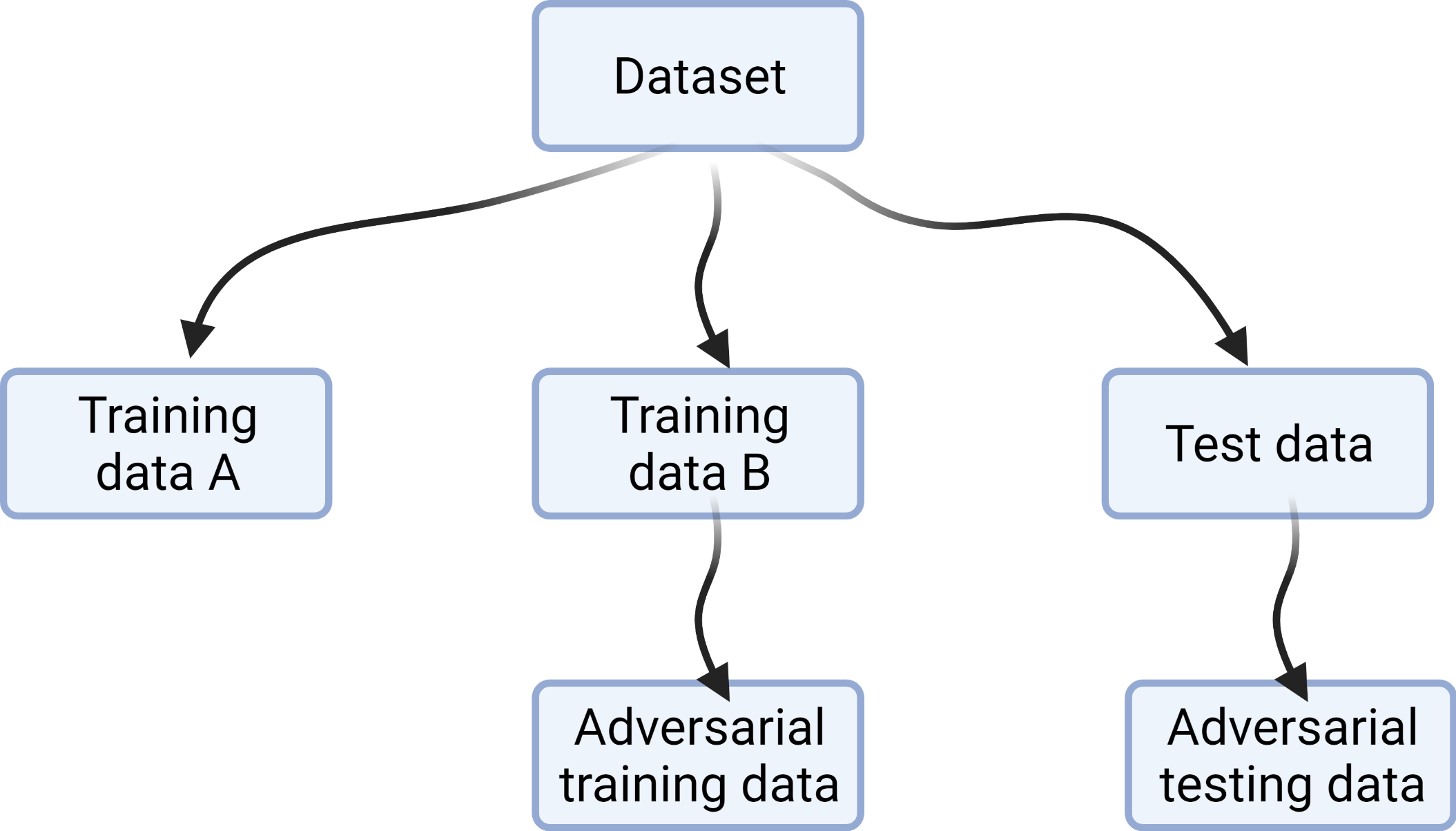}
	\caption[]{The partitioning of the dataset for training and testing of IDS}
	\label{fig:dataset}
\end{figure}

The dataset contains 41 network traffic characteristics, covering three aspects: basic characteristics, content characteristics, and traffic characteristics.  Many attacks are covered in this dataset; they can be further classified into four families of attacks: denial-of-service (DoS) attacks, probe attacks (Probe), root-to-local (R2L) attacks, and user-to-root (U2R) attacks. We use KDDTrain+ in our experiments by dividing it into 80\% and 20\% for training and test data respectively. The training data is divided into two almost equal parts A and B to train the DNN-based IDS separately from other ML-based IDSs so as to examine the transferability property. The data used in the experimental part are summarized in Table \ref{tab:classdistro} and their partitioning is illustrated in Figure \ref{fig:dataset} .

\begin{table}[!h]
	\centering
	\caption{Summary of the network traffic dataset.\label{tab:classdistro}}
	\begin{tabular}{cccccc}
		\hline
		& \textbf{Normal} & \textbf{DoS}   & \textbf{Probe} & \textbf{R2L}  & \textbf{U2R} \\ \hline
		\textbf{Training data A}  & 26938  & 18371 & 4663 & 398  & 21  \\ \hline
		\textbf{Training data B}  & 26937  & 18371 & 4662 & 398  & 21  \\ \hline
		\textbf{Test data }  & 13468   & 9185  & 2331  & 199 & 10 \\ \hline
	\end{tabular}
\end{table}

\subsection{Preprocessing}

The network traffic included in this dataset is heterogeneous and contains both numerical and categorical values. Many machine learning algorithms do not support categorical values, hence the need for the numericalization step that transforms these categorical inputs into numerical values. In the case of the NSL-KDD dataset, the categorical features are "flag", "protocol type" and "service".   Another important aspect is feature scaling, which consists of converting all features to the same scale to ensure that all features contribute equally to the result and also to help the gradient-based ML algorithms converge faster to the minima. We restrict our study to a binary classification where we consider any type of attack as "intrusion" and the rest as "normal" traffic. 
\subsection{Building Anomaly-based Intrusion Detection Systems}
%

TensorFlow \cite{tensorflow2015-whitepaper} is used to build the DNN-based IDS, it consists of two hidden layers with 512 units each. As an activation function, we use Rectified Linear Unit (ReLU) to increase the non-linearity. To prevent overfitting, a dropout layer with a 20\% dropout rate is placed after each hidden layer. ADAM and categorical cross-entropy are used as an optimization algorithm and loss function respectively. In the end, the logits are converted to probabilities using a softmax layer. The final prediction is assigned to the highest probability class.

We acknowledge the use of Scikit-learn \cite{scikit-learn} to build five ML-based IDSs. The default settings were maintained. These five ML algorithms were selected due to their popularity in the ML community: Support Vector Machines (SVM), Decision Tree (DT), Logistic Regression (LR), Random Forest (RF), and Linear Discriminant Analysis (LDA). We also construct an ensemble IDS by grouping these five ML algorithms where the final prediction is made using the majority voting rule.

\subsection{Transferability of Adversarial Attack in Black-box Settings}

In order to test the transferability property of adversarial attacks, we build a DNN-based IDS where we generate adversarial network traffic records in a white-box setting and then test them against five different ML-based IDSs. Note that the five ML-based IDSs are trained on a different dataset (Training data B) and the adversary records were generated without assuming any knowledge of the internal architecture of these five ML-based IDSs, which means that we are working under a black-box setting assumption.

Two adversarial attacks were implemented to generate adversarial network traffic records: FGSM and PDG. For this, we use Adversarial Robustness Toolbox (ART) \cite{nicolae2018adversarial}. The experiments are repeated by increasing the attack strength $\epsilon$ to investigate the amount of perturbation required for the adversarial attack to be transferred from the DNN-based IDS to the other five ML-based IDSs in black-box settings.

\subsection{Defenses against the Transferability of Adversarial Attacks}

Since the ensemble technique is known to increase accuracy over a single classifier \cite{rokach2010ensemble}, we want to examine whether it can also increase its robustness against the transferability of adversarial attacks in a black-box setting. To do so, we construct an ensemble IDS based on the previous five ML-based IDSs and use the majority voting rule to obtain the final decision. 

The second defense we consider is the Detect \& Reject method \cite{grosse2017statistical}, which involves training our IDSs to detect not only "abnormal" and "normal" traffic, but also a third class called "adversarial". Thus, whenever the IDS decides that a network traffic record is adversarial, it is rejected. We implement this method on the five ML-based IDSs and examine their robustness to the adversarial attack transferability property.

\section{Experimental Results}
\label{sec:er}
In this section, we present the results of experimenting our appraoch. Subsection (A) illustrates the effect of the transferability property of adversarial attacks on the five ML-based IDSs.  In subsection (B), we examine the robustness of the ensemble IDS against these attacks in black-box settings. Subsection (C) illustrates the robustness improvement of all IDSs after adding the detection and rejection mechanism to the five ML-based IDSs. 

\subsection{Transferability of Adversarial Attacks in Black-box Settings}

\begin{figure*}[!ht] 
	\centering
	\subfigure[ Transferability from DNN-based IDS to SVM-based IDS]{%
		\centering
		\includegraphics[width=0.45\textwidth]{"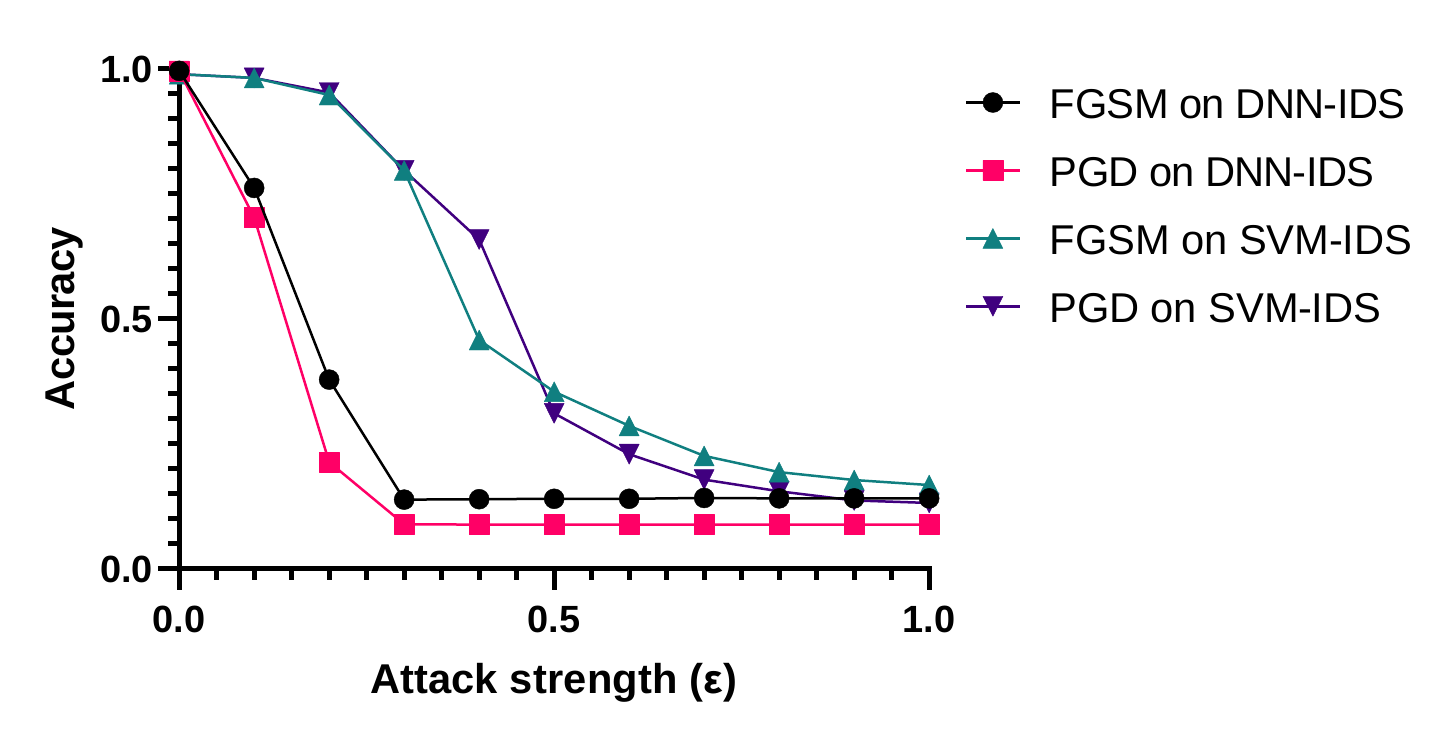"}%
		\label{fig:SVM}%
	}\hfil
	\subfigure[ Transferability from DNN-based IDS to DT-based IDS]{%
		\centering
		\includegraphics[width=0.45\textwidth]{"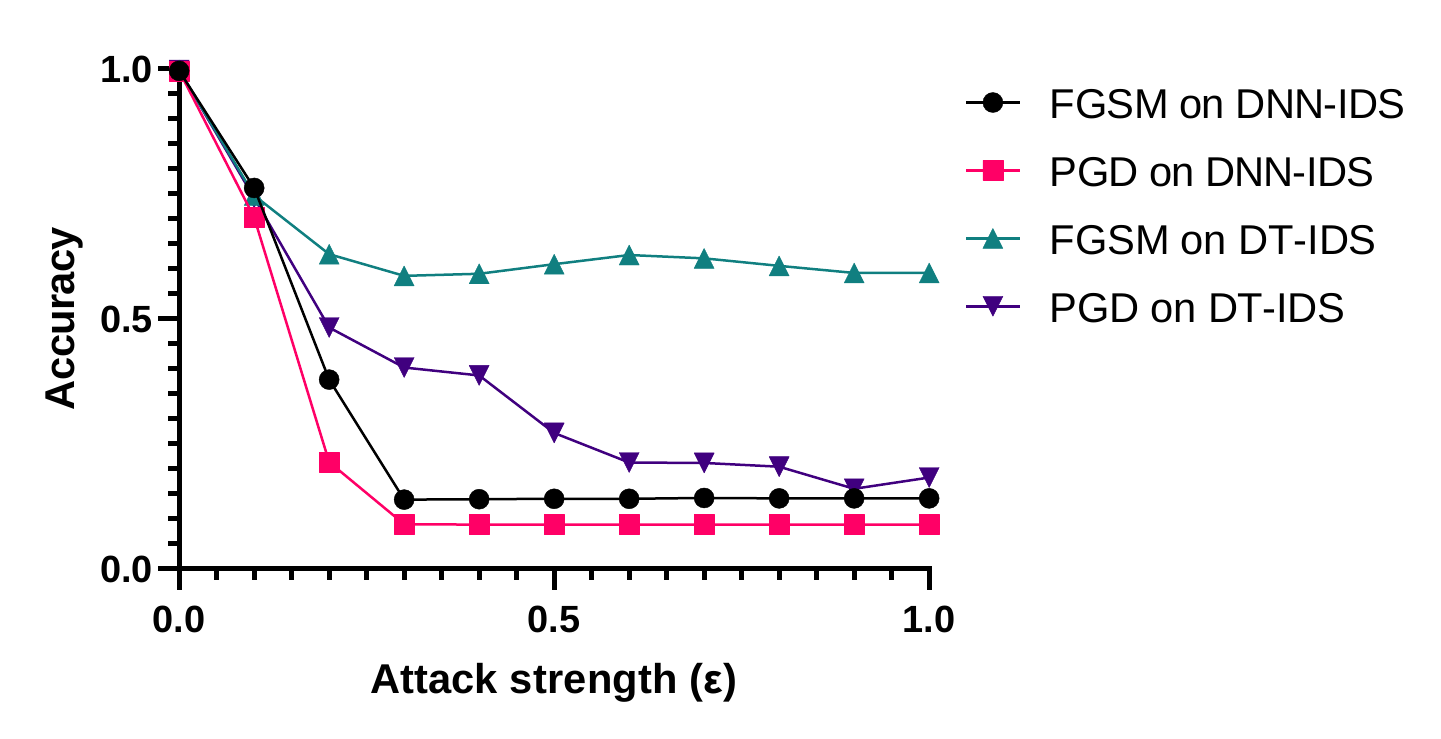"}%
		\label{fig:DT}%
	}
	
	\subfigure[ Transferability from DNN-based IDS to LR-based IDS]{%
		\centering
		\includegraphics[width=0.45\textwidth]{"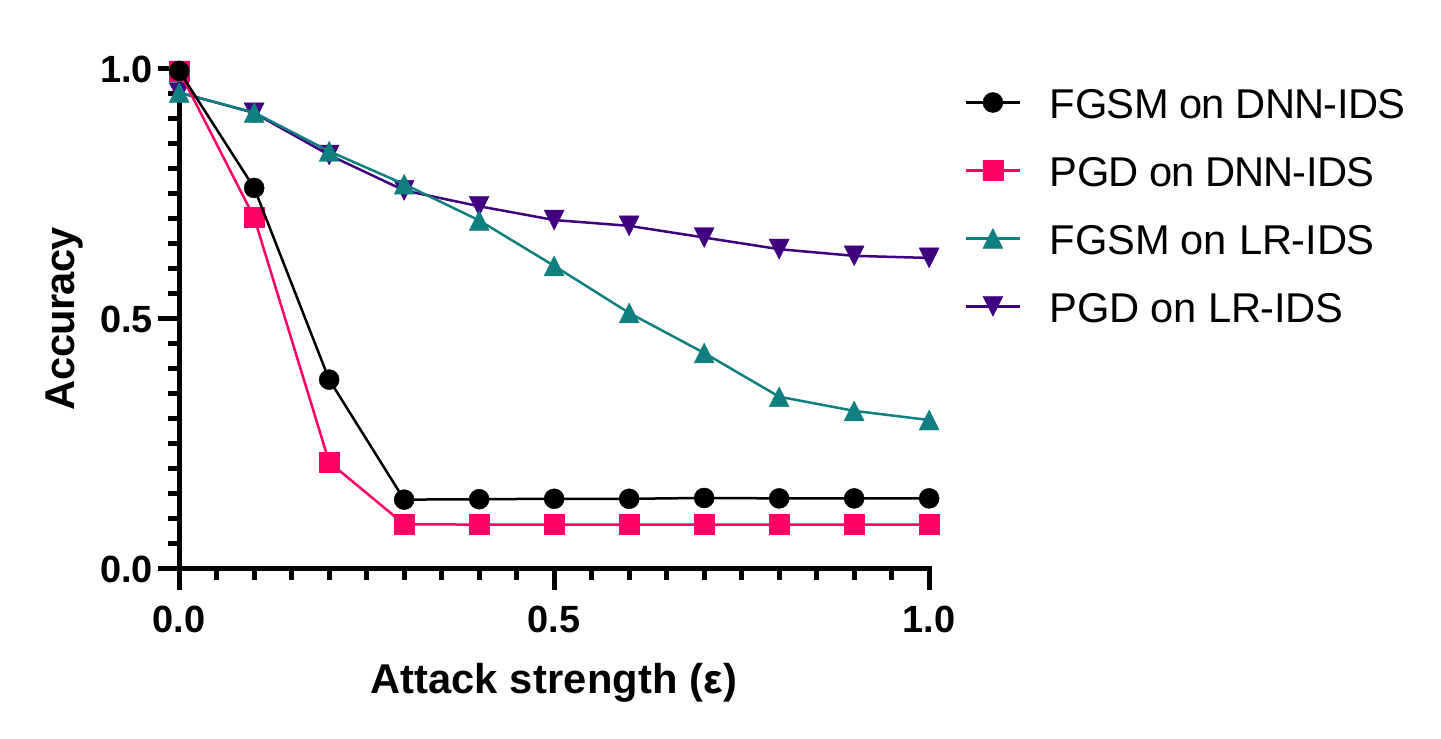"}%
		\label{fig:LR}%
	}\hfil
	\subfigure[ Transferability from DNN-based IDS to RF-based IDS]{%
		\centering
		\includegraphics[width=0.45\textwidth]{"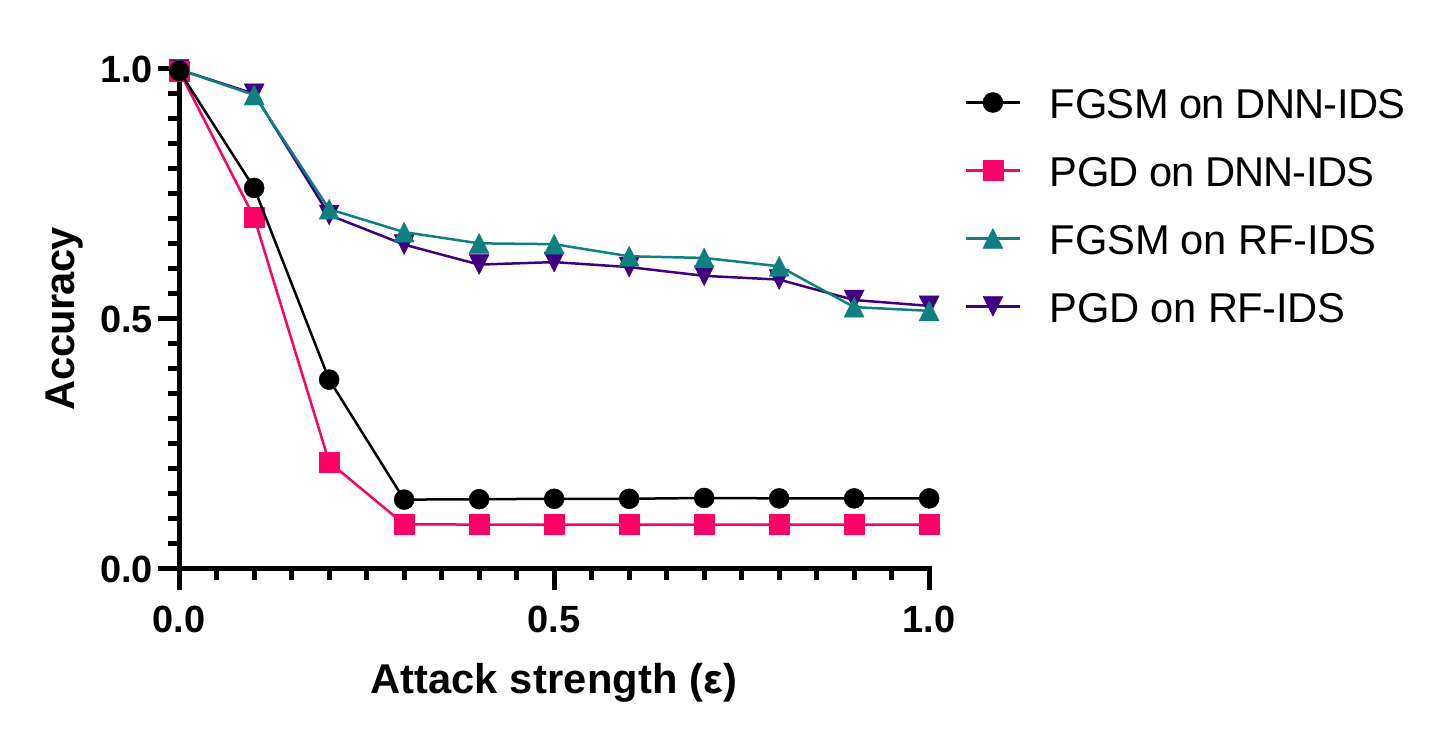"}%
		\label{fig:RF}%
	}
	\subfigure[ Transferability from DNN-based IDS to LDA-based IDS]{%
		\centering
		\includegraphics[width=0.45\textwidth]{"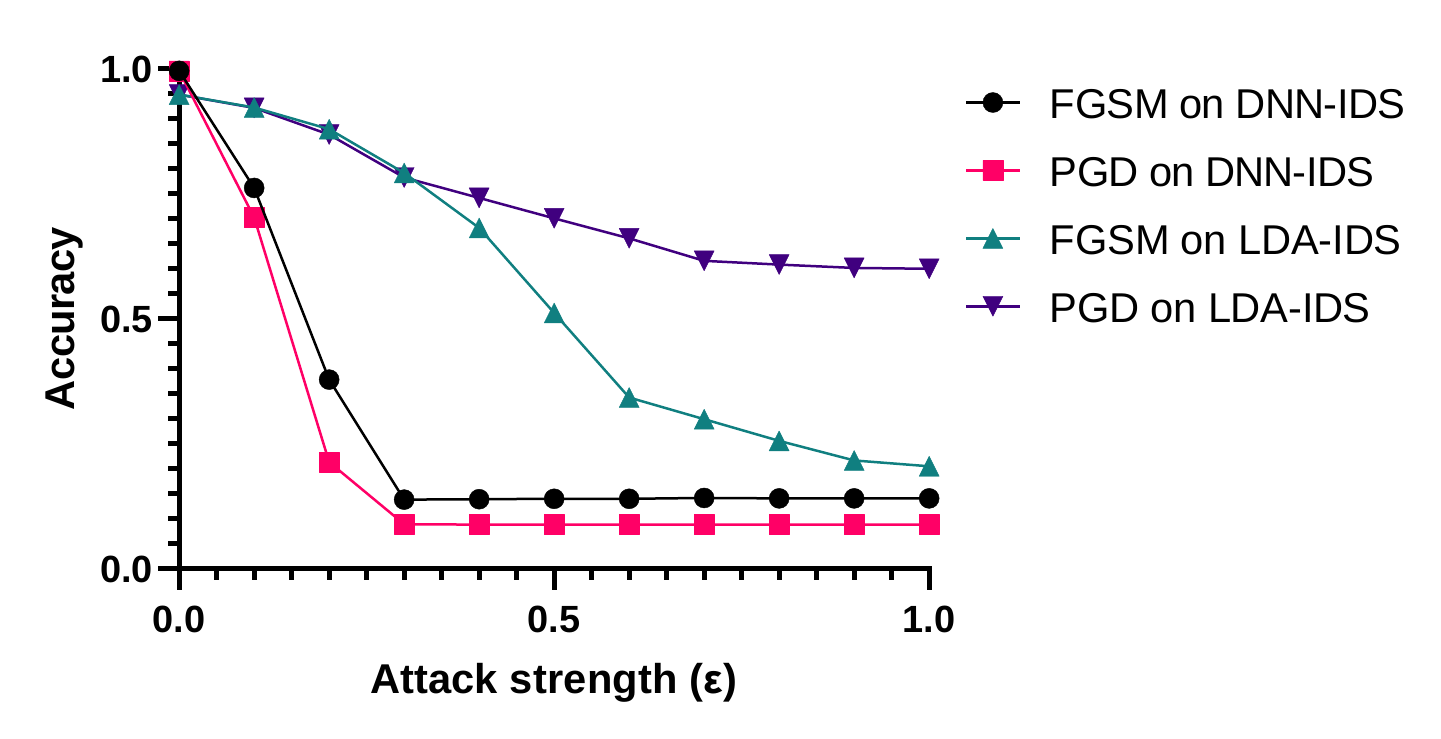"}%
		\label{fig:LDA}%
	}
	
	\caption{Transferability of adversarial attacks against ML-based intrusion detection systems in black-box settings}
	\label{fig:Transf to ML}
\end{figure*}

In this study, we use two adversarial attacks: FGSM and PGD to generate adversarial network traffic records from "Test data". These adversarial records are specifically designed to fool DNN-based IDSs since both attacks have access to the internal architecture of DNNs. As mentioned earlier, we train the DNN-based IDS using "Training data~A", while the other 5 ML-based IDSs are trained using "Training data B". Adversarial traffic records are used at test time to attempt to mislead the IDSs. As shown in Figure \ref{fig:Transf to ML}, increasing the attack strength ($\epsilon$) further degrades the accuracy of the DNN-based IDS.  When testing these adversarial network traffic records on the five ML-based IDSs, we find that their accuracy decreases, even though the attacks do not have access to their internal architectures. We also note that although the accuracy of the ML-based IDSs did not deteriorate as much as the DNN-based IDSs, some models were more vulnerable than others. This may be due to their differentiability property, i.e., they are composed of differentiable elements, since the decision tree and the random forest, whose accuracies were least affected, are non-differentiable models that are not amenable to gradient descent due to their Boolean nature, unlike SVM or logistic regression for example.

\subsection{Ensemble Intrusion Detection System Robustness} 

\begin{figure}[tbh!]
	\centering
	\includegraphics[width=1\linewidth]{"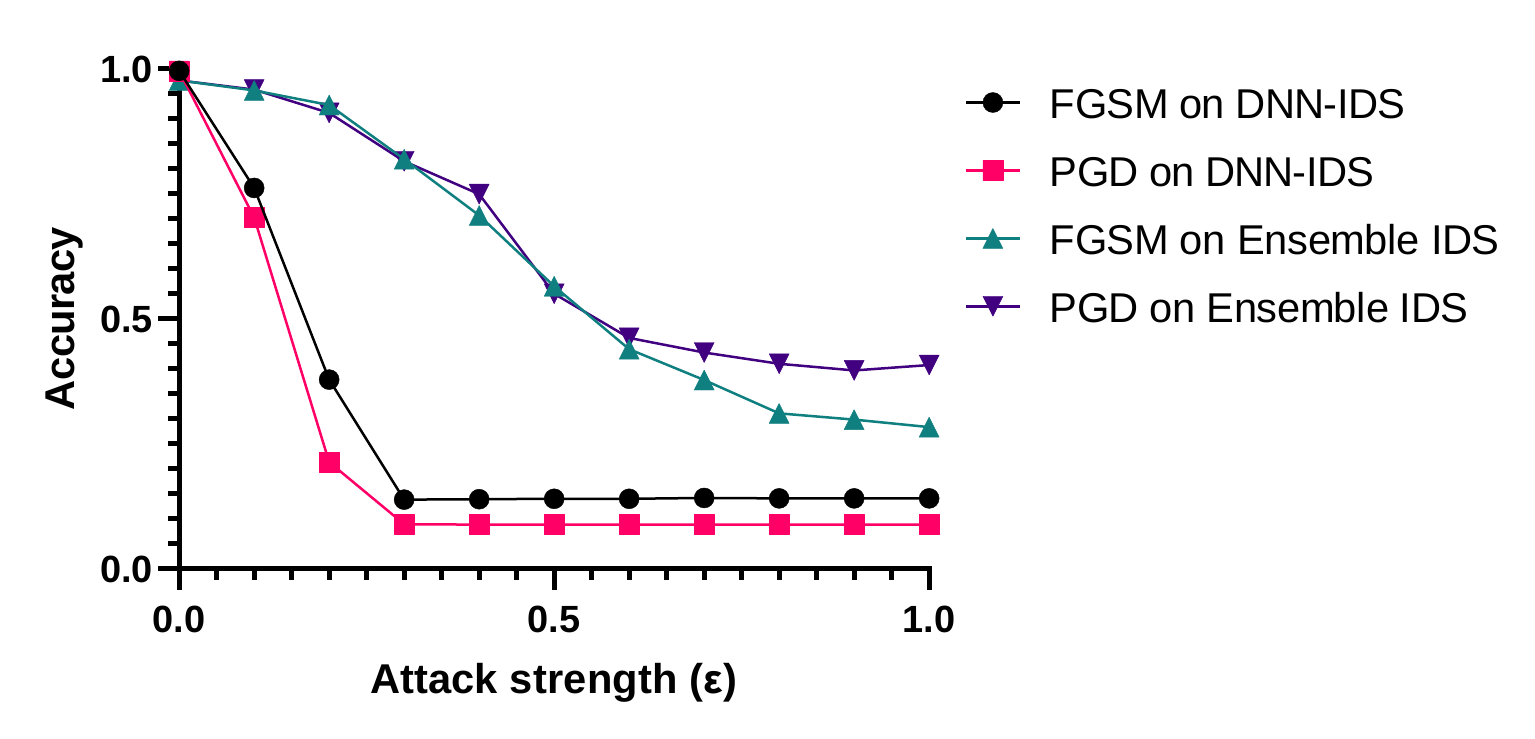"}
	\caption{ Adversarial attack transferability from DNN-based IDS to Ensemble IDS}
	\label{fig:transf to Ens}
\end{figure}

Since the ensemble technique is known to improve accuracy over a single model, we investigate whether it could also improve robustness. To this end, we construct an ensemble IDS based on the five ML-based IDSs using the majority voting rule. The same setup as in the previous experiment is maintained, which means that the 5 ML models used to build the ensemble model are trained using the "Training data B". The adversarial traffic records are generated from the "Test data" using the FGSM and PGD attacks. These adversarial records are designed to fool DNN-based IDS since both attacks can only access the internal architecture of the DNN model. As shown in Figure \ref{fig:transf to Ens}, the ensemble IDS is not able to resist the transferability property of adversarial attacks, even though no information about the ensemble IDS was used to generate these adversarial records. This shows the ease of an evasion attack against an intrusion detection system without even knowing its internal architecture, simply by building a surrogate IDS (DNN-based IDS in our case) and generating adversarial network traffic for this surrogate model. 

In practice, if we consider malicious network traffic, such as HTTP traffic that seeks to connect to dangerous URLs, like command-and-control [C\&C] servers, an attacker could use adversarial attack techniques to disguise this malicious network traffic as normal traffic for the intrusion detection system while retaining its malicious aspects. This can be done by adding small amounts of specially designed data to that network traffic as a padding for example. Therefore, the attacker could evade the intrusion detection system.

\subsection{Detect \& Reject for Adversarial Network Traffic} 

\begin{figure*}[tbh!] 
	\centering
	\subfigure[ Detect \& Reject for SVM-based IDS]{%
		\centering
		\includegraphics[width=0.5\textwidth]{"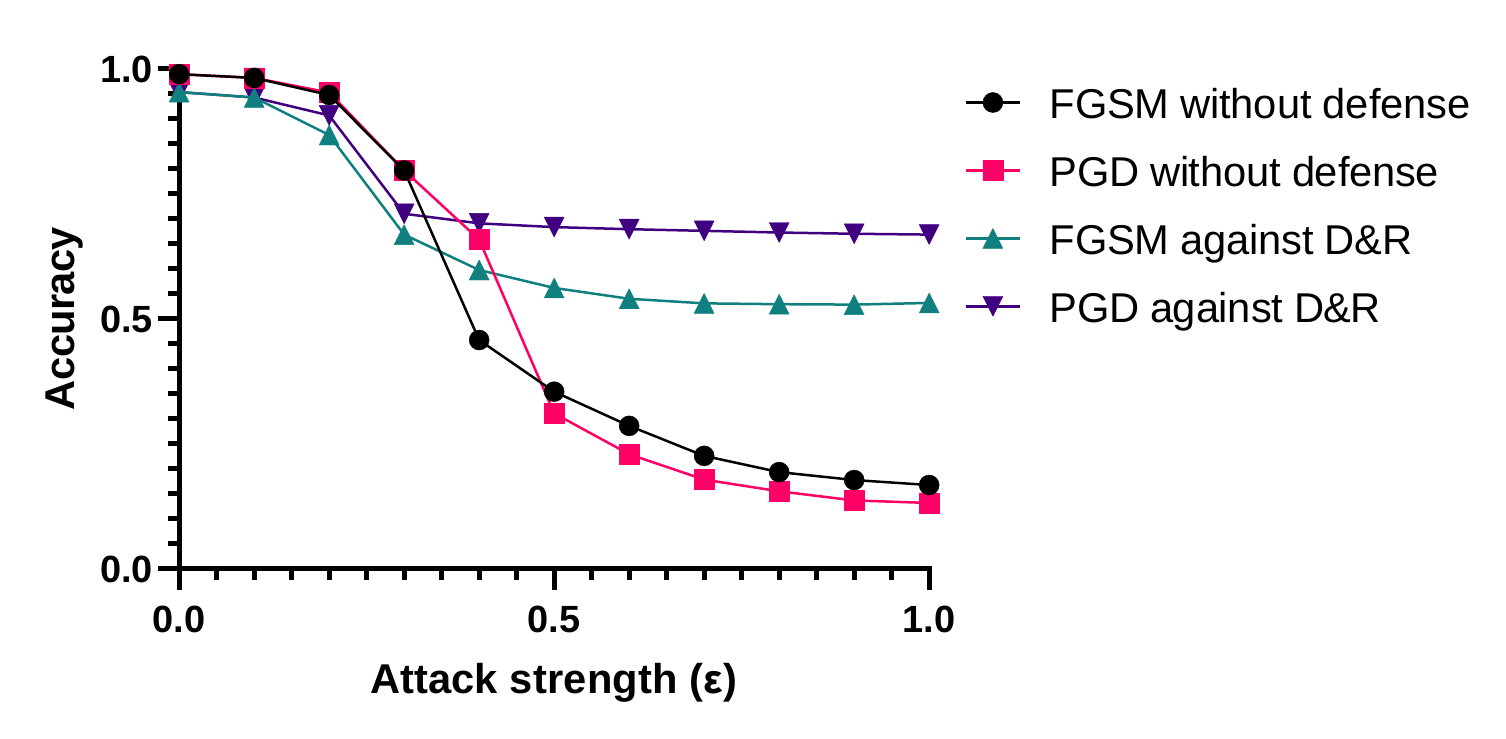"}%
		\label{fig:DKF SVM}%
	}\hfil
	\subfigure[Detect \& Reject for DT-based IDS]{%
		\centering
		\includegraphics[width=0.5\textwidth]{"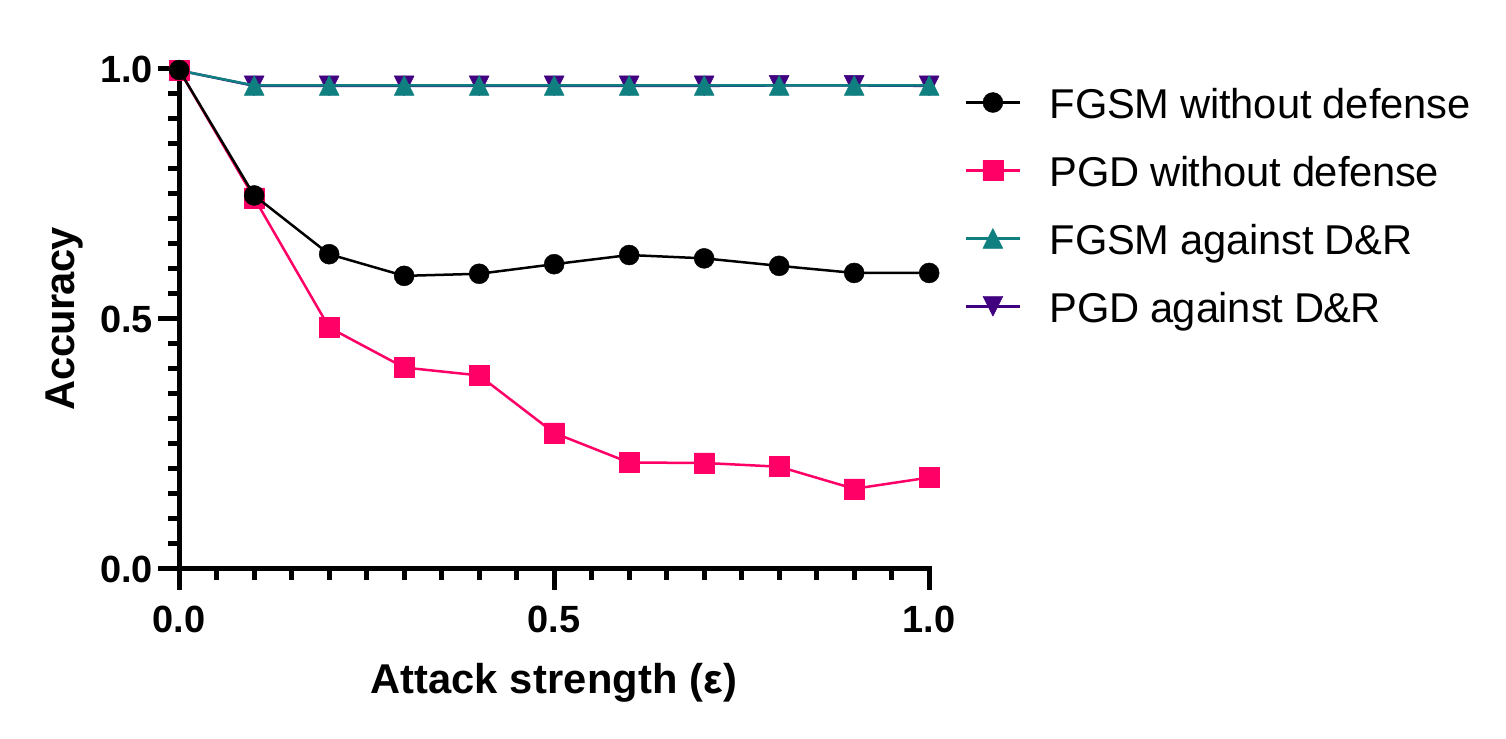"}%
		\label{fig:DKF DT}%
	}
	
	\subfigure[Detect \& Reject for LR-based IDS]{%
		\centering
		\includegraphics[width=0.5\textwidth]{"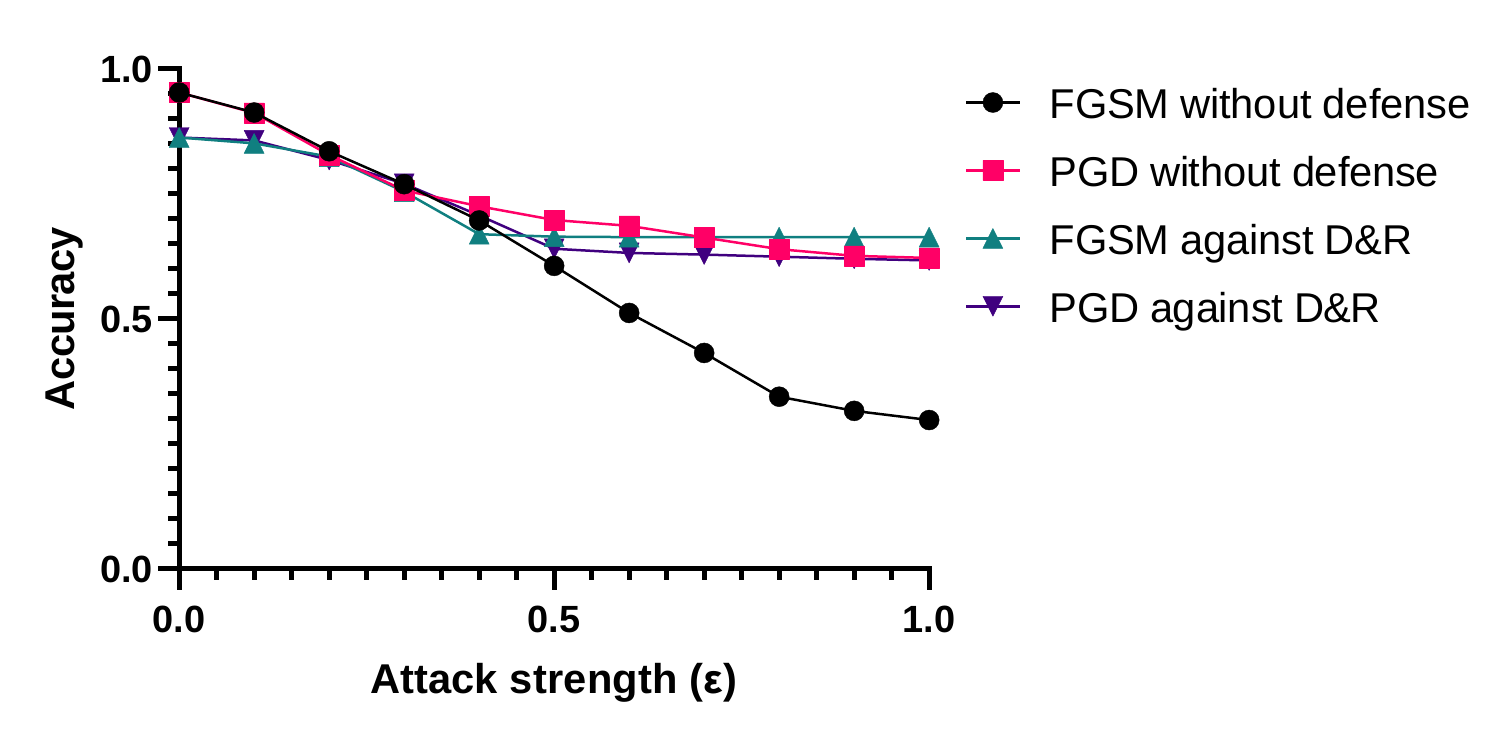"}%
		\label{fig:DKF LR}%
	}\hfil
	\subfigure[ Detect \& Reject for RF-based IDS]{%
		\centering
		\includegraphics[width=0.5\textwidth]{"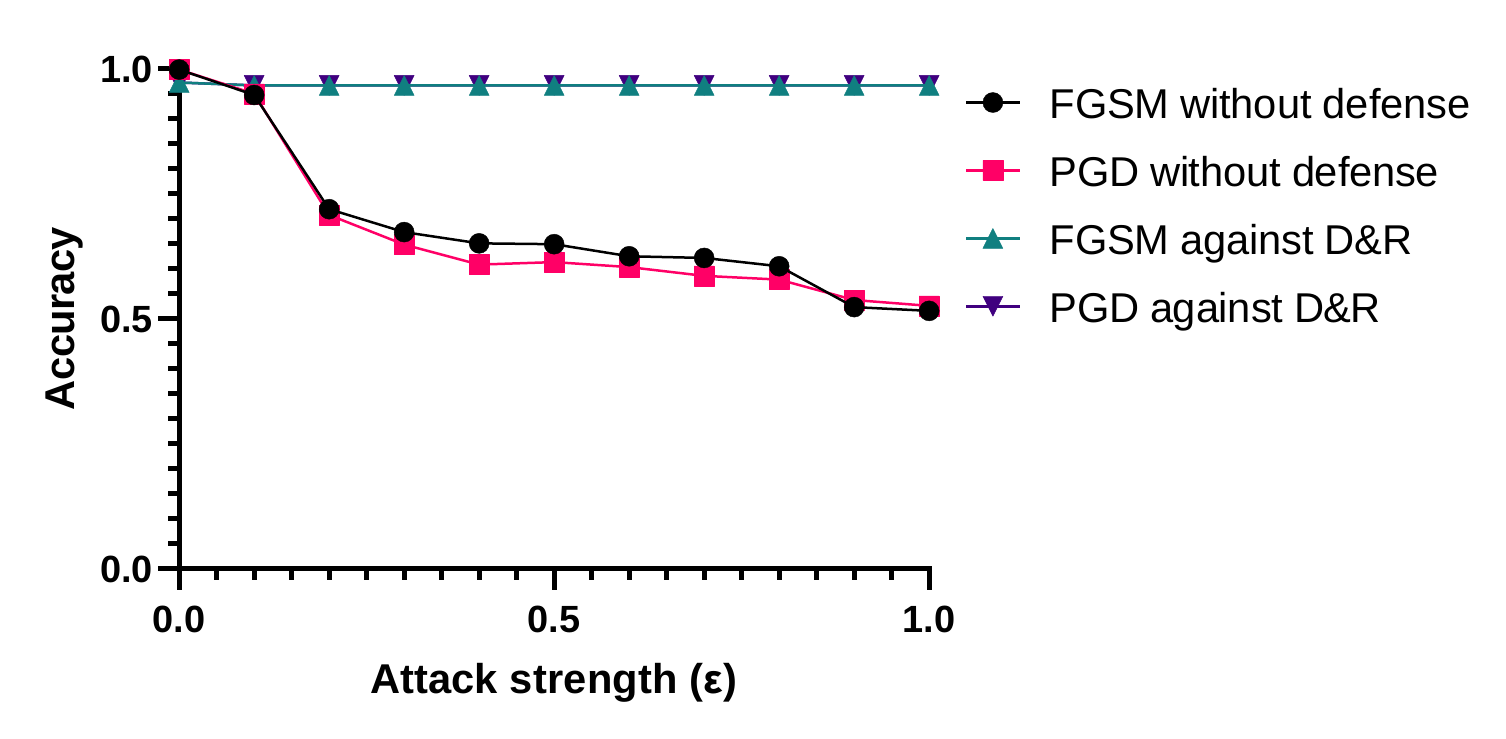"}%
		\label{fig:DKF RF}%
	}
	\subfigure[ Detect \& Reject for LDA-based IDS]{%
		\centering
		\includegraphics[width=0.5\textwidth]{"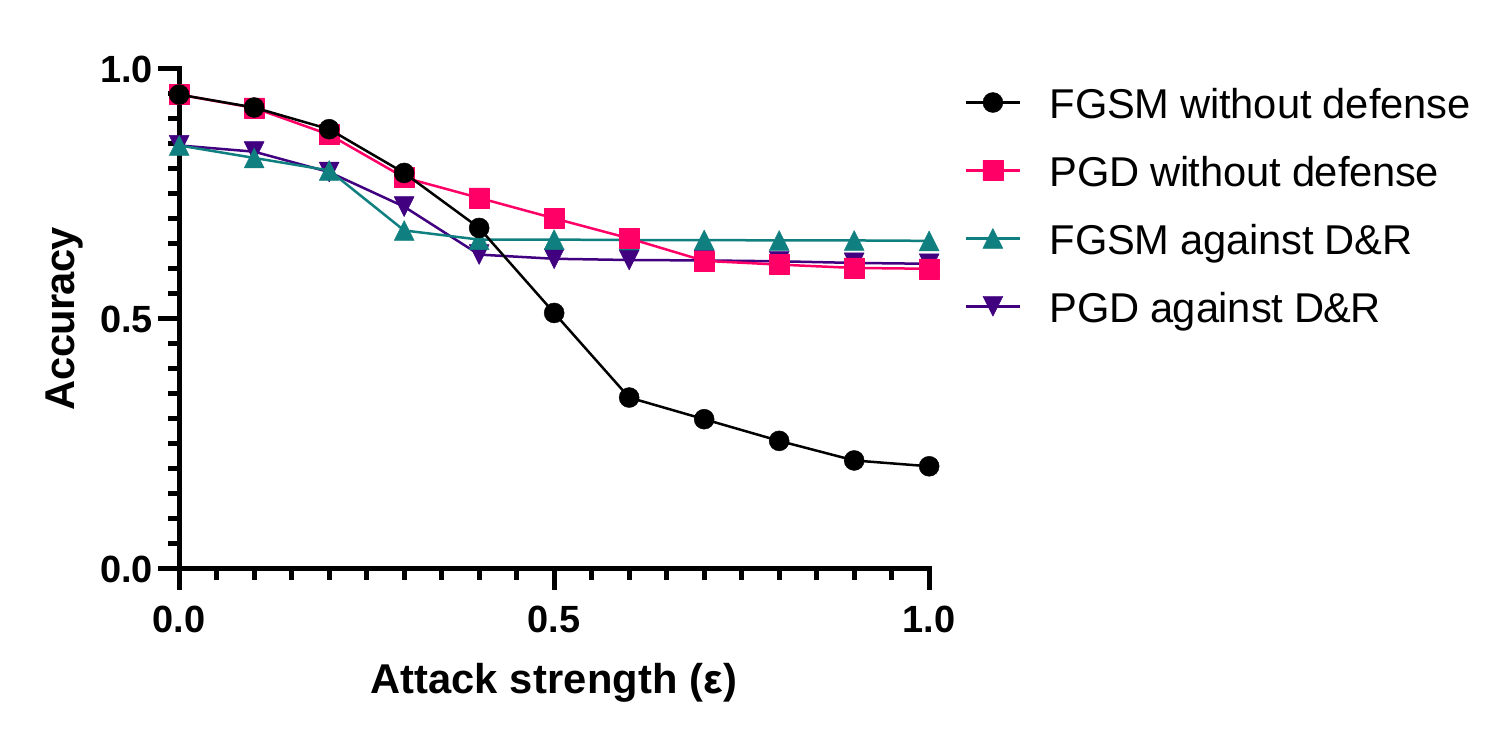"}%
		\label{fig:DKF LDA}%
	}
	
	\caption{ Detect \& Reject as a defense against the transferability property of adversarial network traffic }
	\label{fig:Transf to DKF ML}
\end{figure*}

In order to limit the effect of adversarial attacks in a black-box context, we implement the Detect \& Reject method in each of the five ML-based IDSs. This method consists of re-training the model to detect not only "abnormal" and "normal" traffic but also "adversarial" traffic. PGD is used against "Training data B" to generate "Adversarial data". After that, the ML model uses a combination of "Training data B" and "Adversarial data"  during the training phase to learn to distinguish the three classes. During the prediction phase, any network traffic record recognised as "adversarial" will be rejected. As shown in Figure \ref{fig:Transf to DKF ML}, all five ML-based IDSs have improved their robustness against adversarial attacks. Decision Tree and Random Forrest, which is an ensemble version of Decision Tree, have the highest detection rates of adversarial network traffic compared to the other IDSs.

\section{Conclusion and Future Work}
\label{sec:cfw}

From an intrusion detection system perspective, adversarial attacks are a serious threat, as a small intentional perturbation of network traffic can mislead the system. To generate these adversarial records, the attacker must have access to the internal architecture of the machine learning model. However, by exploiting the transferability property of adversarial attacks, he can mislead other intrusion detection systems without having any knowledge about them. Ensemble IDSs, although known to improve model accuracy, are vulnerable to these attacks and thus cannot improve model robustness. On the other hand, Detect \& Reject has shown through our experiments to be a suitable built-in defense for intrusion detection systems against adversarial attacks.
An interesting future work would be to design more effective defenses to limit the effect of adversarial attacks against intrusion detection systems.

%
%
\bibliography{BibTexDataBase.bib}

\bibliographystyle{acm}

\end{document}